Grant-Making Under Uncertainty:

Response to the NIH Science of Science Working Group's Request about Optimizing NIH Decision-Making Processes

Daniel L. Goroff, February 2026[1]

*"Perhaps the most important ideas of all are meta-ideas — ideas about how to support the production and transmission of other ideas... Human history teaches us that economic growth springs from better recipes, not just from more cooking."*

-Nobel laureate Paul Romer[2]

## 0. Introduction

The NIH spends well over $30 billion annually on extramural research grants.[3] That is, on Paul Romer's view quoted above, lots of cooking. Many who cherish NIH and the work it supports simply want more. This Working Group is, however, concerned with better recipes. We are scarcely alone in that. It is not hard to find food critics with ideas about how they would run an eating establishment better. And it is especially easy to find such critics who, because they lack the conceptual framework of a restauranteur, offer interesting and insightful opinions that are nevertheless difficult to evaluate, prioritize, or implement.

The "Science of Science" is here to help. It is the academic field that conducts rigorous research on new recipes to, as Romer puts it, "support the production and transmission of other ideas." Similarly, "Metascience" is a more recent term insofar as it refers to sprawling and multidisciplinary takes on similar questions. The "Economics of Science and Technology" has a long, distinguished, and disciplined history that draws on many economic subfields ranging from industrial organizational, innovation theory, and mechanism design to growth theory, labor economics, and decision analysis. The last of these, particularly the study of decision-making under uncertainty, has particular relevance to the redesign of federal funding procedures. It can, as we shall see here, give precise meaning and clear usefulness to terms that often appear casually such as "optimization," "theories of change," "trade-offs," and "risk."

---

[1] Written for the NIH's Council of Council Working Group on the Science of Science in response to a request to members for perspectives on "optimizing" NIH grantmaking processes. Opinions expressed here are not necessarily those of the Alfred P. Sloan Foundation or of any of the author's other institutional affiliations.
[2] The line quoted in the epigraph is from Paul M. Romer, "Economic Growth," in David R. Henderson, ed., The Concise Encyclopedia of Economics (Library of Economics and Liberty, 2008). For his scholarly treatment of "meta-ideas," see Paul M. Romer, "Two Strategies for Economic Development: Using Ideas and Producing Ideas," World Bank Economic Review 6, suppl. 1 (1992): 63–91, https://doi.org/10.1093/wber/6.suppl_1.63.
[3] See National Institutes of Health, "Budget," Office of Budget, https://www.nih.gov/about-nih/organization/budget, and the NIH Data Book, https://report.nih.gov/nihdatabook/ (extramural obligations were $34.9 billion of a $47.7 billion appropriation in FY2023; the total NIH budget was approximately $30 billion as recently as FY2015).

To expand on just one of these examples for now, critics like to complain that science funders do not take enough risk. What does that mean? Is an option risky because it has a considerable probability of turning out moderately bad or because it has a modest probability of turning out catastrophically bad? Would buying lots of lottery tickets count as the kind of "high risk-high reward" strategy advocated? And since it never repeats, what does it even mean for an event like a scientific discovery to have a particular probability anyway? Recommendations that sound reasonable enough can have unintended consequences unless examined carefully in the context of a more comprehensive framework.

In fact, NIH not only receives lots of untethered recommendations like this, it invites them. Rather than explicitly "optimizing" or deliberately designing their research investment decisions, most science funders like NIH traditionally outsource some of their responsibility by giving panels of outside experts a number of somewhat related proposals to read, then asking them directly which *they* would fund. This venerable process, called peer review, has many advantages—including that decisions can be routinely made relatively free from political influence. Indeed, peer review has served NIH and the United States quite well for many decades.[4] But to further improve NIH's recipe for cooking up and distributing resources, unstructured advice is not enough. As an organizing device, it often helps to start with first principles.

## 1. Basics of Decision-Making

The first principle of science funding is that it is a particular kind of decision-making under uncertainty. From that point of view, allocating university grants to fund research has much in common with allocating financial investments to fund retirement. Most humans are not very good at this, and it is not easy to evaluate who is and who is not without getting into complex counterfactuals. On the other hand, there are rather well-developed ways to structure and carry out decision-making under uncertainty that are, oddly enough, almost universally ignored in practice but provably optimal in theory. How close can we come to applying lessons from Decision Science at NIH?

[4] Danielle Li and Leila Agha, "Big Names or Big Ideas: Do Peer-Review Panels Select the Best Science Proposals?," Science 348, no. 6233 (2015): 434–438, https://doi.org/10.1126/science.aaa0185, finding that NIH study-section scores predict subsequent publications, citations, and patents.

The answer depends on how NIH is evaluated. A key insight about assessing decisions made under uncertainty is not to look at the outcomes only, but rather at the quality of the process by which the decision was made, i.e., at the recipes. Robert Rubin is just one of many successful financiers who have made a point of this.[5] After all, a decision-maker could place a very good bet, but the dice just happen to roll unfavorably when it counted. Or such a decision-maker could have made a very bad bet but then just happened to come up lucky. Outcomes do matter, of course. In the long run, though, a high-quality process will deliver the best overall returns.

The next section discusses the shape that such a high-quality process can and should take. A sufficiently general form of utility maximization is the short answer. And that conclusion is based on theorems, not just on opinion or tradition.

## 2. Optimization

Optimization usually refers to finding the maximum value of some objective function defined on a space of possible inputs. Deep Learning algorithms have popularized the idea of exploring that space guided by the gradient of an objective function, for example. That's fine if you know the space along with the function and its derivatives. It's hardly that simple in the NIH setting. The consequences of funding a research project are not certain in advance. That is why it is called research. After all, if we knew just if and how a given experiment would turn out, we could support it by writing a contract rather than making a grant.

Under mild rationality assumptions as described in (2a) below, decision-making under uncertainty can always be described as maximizing the expected value of a utility function. Moreover, any decision rule that does not admit such a representation is necessarily suboptimal (2b). And even if subjective, both the utilities of complex (e.g., non-monetary) outcomes as well as the probabilities of complex "states of the world" can, in principle, be systematically elicited and calibrated against canonical probabilities that do have objective values (2c).[6] The important lesson is that it is always possible, and, as a practical matter, advisable to distinguish between estimating the *value* of a given outcome and estimating *beliefs* about the probability that outcome will occur. Conflating these two tasks is common, confusing, and can be avoided by assigning them to different parties (as NIH effectively does, as described in §3, without necessarily using the results this way).

---

[5] See Jonathan Baron and John C. Hershey, "Outcome Bias in Decision Evaluation," Journal of Personality and Social Psychology 54, no. 4 (1988): 569–579, https://doi.org/10.1037/0022-3514.54.4.569, on the tendency to judge a decision's quality by its result rather than by the information and reasoning available when it was made.

2a. **Representation theorems about optimization:** Any talk about optimizing something first requires some kind of consistent ordering. In decision-making under uncertainty, we can imagine there are some underlying preferences over the options available—which are called "acts" in the literature but that are better thought of as research proposals in this context. Specifically, suppose there is a finite set of possible states of the world $S$, and each act (i.e., research proposal) is a random variable $x$ specifies what outcome $x(s)$ you would receive in each state $s$.

In breakthrough research, Leonard Savage[7] starts by asking about what your preferences over such acts look like if they are to be internally coherent? He proposes a small set of consistency conditions including completeness, transitivity, and, most importantly, the "Sure-Thing Principle." (It simply says that if a decision-maker would want option $x$ knowing that event $E$ occurs and also in the case that $E$ does not occur, then the same $x$ should also be preferred if he or she knows nothing about $E$.) Under those assumptions, exist (i) a function $u$ assigning a real number $u(c)$ to each possible consequence $c$, capturing how desirable that consequence is to you, and (ii) a probability distribution $P$ over the states $S$, such that the decision-maker prefers $x$ over $y$ if and only if the expected value of the random variable $u(x)$ with respect to $P$ is greater than the corresponding expected value of $u(y)$. Savage's Representation Theorem is therefore a powerful result. It says that, if you want your decisions under uncertainty to be internally consistent—to respect stable tradeoffs across states and to obey the Sure-Thing Principle—then your choices must take the form of expected-utility maximization.

---

[6] The reference-lottery elicitation procedure described here follows Carl S. Spetzler and Carl-Axel S. Staël von Holstein, "Probability Encoding in Decision Analysis," Management Science 22, no. 3 (1975): 340–358, https://doi.org/10.1287/mnsc.22.3.340.

[7] Leonard J. Savage, The Foundations of Statistics (New York: John Wiley & Sons, 1954; 2nd rev. ed., New York: Dover, 1972). For a retrospective assessment, see Glenn Shafer, "Savage Revisited," Statistical Science 1, no. 4 (1986): 463–485, https://doi.org/10.1214/ss/1177013518.

**2b. This representation avoids non-optimal decision rules**. A procedure for choosing among acts is called a decision rule. As above, suppose there is a finite set of states $S$ and a finite set of actions $A$. A utility function $u$ that assigns the number $u(x(s))$ to the outcome produced by action $x$ in each state $s$. A decision rule is called inadmissible if there exists another rule that yields as much utility in every state and strictly more in some state. Wald's Completeness Theorem[8] says that every admissible decision rule maximizes expected utility with respect to some probability distribution over states (or is a limit of such rules). Equivalently, if you do not choose as if maximizing expected utility for some beliefs, then there is another strategy that makes you as well off or better no matter which state turns out to be true.

Thus, Wald starts from a performance criterion—avoiding being uniformly outperformed—and shows that only expected-utility–maximizing rules survive that test. Savage starts from coherence of preferences and shows that rational choice must take the form of expected-utility maximization. Either way, you cannot avoid expected utility considerations if you want to make optimize the decisions you make under uncertainty.

**2c. Calibrating subjective probabilities and utilities**. The word "subjective" before the words like "utility" or "probability" might suggest that these are arbitrary constructs pulled out of the air. On the contrary, for a given decision-maker they can be systematically elicited and calibrated making use of the canonical probabilities everyone agrees upon for objectively repeatable events like roulette wheel spins. That exercise, as described here, first requires setting references outcomes $B$ and $W$ that are better and worse, respectively, than other possible outcomes. For example, $B$ might be a discovery that lengthens everyone's QALYs by 20% or more and might be a discovery that decreases everyone's QALYs by 20% or more.

Given such a $B$ and $W$, one can use objective (canonical) lotteries as measuring rods to calibrate both utilities and probabilities. To elicit utilities, consider an intermediate consequence $c$. Ask for the objective probability $p$ that would make you indifferent between receiving $c$ for sure and a lottery that yields $B$ with probability $p$ and $W$ with probability $(1-p)$. If your preferences satisfy the usual expected-utility conditions, that indifference implies

$$u(c) = p\,u(B) + (1-p)\,u(W).$$

[8] Abraham Wald, Statistical Decision Functions (New York: John Wiley & Sons, 1950). For an accessible modern exposition of the complete-class theorem, see James O. Berger, Statistical Decision Theory and Bayesian Analysis, 2nd ed. (New York: Springer, 1985), https://doi.org/10.1007/978-1-4757-4286-2.

Normalizing $u(B) = 1$ and $u(W) = 0$, the utility of $c$ is simply $p$. In this way, canonical probabilities—like the known odds of a fair coin or roulette wheel—serve as a scale for measuring the relative desirability of outcomes.

Similarly, subjective probabilities for events can be calibrated using the same reference outcomes. Let $E$ be an uncertain event (e.g., whether a certain experiment produces a highly cited paper). Ask for the objective probability $q$ that would make you indifferent between (i) a lottery yielding $B$ with probability $q$ and $W$ otherwise, and (ii) receiving $B$ if and only if event $E$ occurs (and $W$ otherwise). Under the usual assumptions about expected utility, that indifference implies $q = P(E)$, your subjective probability of $E$.

Thus, by anchoring judgments to fixed best and worst outcomes and using objective randomization devices as benchmarks, one can jointly and coherently calibrate both utilities and beliefs through observed indifference tradeoffs. In fact, this way of eliciting utilities and probabilities implies that maximizing expected utility is equivalent to maximizing the probability, given your beliefs and values, of the best outcome as opposed to the worst.

A few further observation about this framework worth noting[9]: (i) Utilities can be assigned this way to outcomes of any sort, not just to numerical, despite the impression otherwise given by the prevalence of utility functions in financial analysis where monetary outcomes are paramount; (ii) Taking the best and worse reference outcomes in terms of QALYs, for example, inextricably brings in from the beginning a concern with health outcomes rather than trying to urge decision-makers to somehow take those into account later; (iii) precision is not so necessary beyond, say, a Likert scale estimate in the sense that, assuming a small number of states of the world, what to select among a small number of options is typically the same for a wide range of probability and utility estimates.

---

[9] Also worth noting are recent generalizations of the framework in which the state space and the outcome space can be subjectively constructed, too, while the acts to be chosen from are each interpreted as an algorithm-like procedure that, when executed, produces particular outcomes given particular states of the world. See Joseph Y. Halpern and Rafael Pass, "Algorithmic Rationality: Game Theory with Costly Computation," Journal of Economic Theory 156 (2015): 246–268, https://doi.org/10.1016/j.jet.2014.04.007, which models acts as algorithms carrying explicit computational costs rather than as simple functions from states to outcomes; and, on endogenously or subjectively constructed state spaces, Joseph Y. Halpern, "Alternative Semantics for Unawareness," Games and Economic Behavior 37, no. 2 (2001): 321–339, https://doi.org/10.1006/game.2000.0832.

## 3. NIH Processes and Decision Analysis

For the sake of simplicity, let's begin by considering a funder whose only task is to select one of several proposals to support that would each cost about the same amount. As a further oversimplification, let's also assume each proposal consists of a plan to test one particular scientific hypothesis.

NIH currently goes through a two-stage process for making such a decision. In many ways, that process mimics the ideal by separating out value judgements from belief judgements as described below. To express this in the language of decision analysis, we begin by looking at a proposal as describing a potential "act" that, if selected for enactment, would yield different "outcomes" depending on which potential "state of the world" is realized by chance.

3a. **What principal investigators present.** Imagine a research proposal to test a scientific hypothesis. The principal investigator presents plans to run an experiment that is designed to offer convincing evidence about that hypothesis. Depending on the findings, the project also plans to yield outcomes such as new papers, patents, trainees or techniques. In advance of conducting the experiment, though, no one knows for certain whether the hypothesis under investigation is true or false. That depends on nature.

More precisely, decision theorists imagine the different "states of the world" that are possible as forming a finite set $S$. Think of each such state in $S$ as represented by one of many rectangles that cover a wall. For the finite list of True/False questions you might ever ask about how the natural world operates, now and forever, each square lists an answer. When you set up an experiment to test a particular hypothesis, you image all the rectangles where that hypothesis is true turning green and all the ones where it is false turning red. There may be more of one color than the other, but we humans can usually only make guesses about that rather than know for sure. When you do run that experiment, the Greek goddess of chance named Tyche comes into the room blindfolded and throws a dart at the wall that is covered with rectangles. Depending on the color where her dart lands, the experimental provides evidence accordingly—though, depending on the experimental design, perhaps imperfectly.

From this point of view, a proposal is a random variable $x: S \to O$ where $S$ denotes the states of the world and $O$ denotes the set of possible outcomes. Think of each member of $O$ as the deliverables produced. More specifically, each element of $O$ is a vector. Each component corresponds to a kind of deliverable, and the entry in that component corresponds to a measure of how much the proposers eventually deliver. So, for example, from the many possible success criteria, you are only interested in seeing highly cited papers, patent applications, and trained postdocs, then (10,000, 4, 2) would represent total citations, the number of patent filings, and the postdoc count that a given proposal could produce. In decision theory, it is customary to refer to the function $x$ as an "act" and to denote the set of all acts under consideration by the letter $A$, but we can think of $x$ as standing for a proposed experiment.

Principal investigators may have lots of hunches about the number of red or green rectangles on Tyche's wall, typically expressed by saying this is a "long shot" or some similar phrase. They may also promise lots of desirable deliverables if the dart lands on a green rectangle. But typically, proposal writers do not say much about the existence, much less the probability of the dart landing on, red ones. Proposers prefer instead to describe all the wonderful outcomes expected—conditional on Tyche hitting a green one.[10] (This is the issue of asymmetric information, i.e., that proposers have little incentive to reveal all they know.

3b. **What study sections present.** Ideally, an NIH study section has the expertise to say much more about probabilities than the principal investigators can or do. Specifically, we want them to provide details about the kinds of states in $S$ and about the probability $p(s)$ they would assign to an $s$ representing a kind of states in $S$. In terms of Tyche's wall, we study sections to take their best guesses both about the distribution of colors, perhaps even beyond just red and green. Some of the rectangles, for example, might be checkered to represent a state of the world in which the experimental results, which themselves depend on some randomness, actually check out properly. In other words, an unchecked, solid green rectangle stands for the return of a false positive and an unchecked, solid red rectangle stands for the return of a false negative.

[10] In other words, the proposal determines an $x$ but the proposers, when promising deliverables, usually write about the conditional expectation $\mathbb{E}[\ x \mid Green]$.

There is also another valuable kind of probabilistic information that the study section could provide, but rarely does very explicitly, concerning how their prior beliefs about the distribution of green or red rectangles change given the results of a proposed experiment. For example, imagine that the experiment proposed by $x$ is run and says the hypothesis is true. How much would that change the study section's posterior beliefs about the distribution of red and green rectangles on Tyche's Wall? (In general, change in belief due to a Bayesian update can be measured by the Kullback–Leibler Divergence.[11] Its expected value, called the Mutual Information, measures how informative you expect an experiment to be. That number could also be added to the vector of promised outputs by which a proposal is then judged by an institute council.)

3c. **What institute councils present.** It is up to NIH officials to say how much they would value various outcomes in the set $O$. Recall that each member of $O$ is a vector of deliverable targets $x(s)$ to be produced by a given proposal $x$ in state of the world $s$. Of particular interest might be measures of the output of papers, patents, people, practices, perspectives, and progress, for example. Each of these six sample objectives can have at least one of its own rating scales, of course. Proposals typically promise that, if the experiment turns out favorably, there will be three highly cited papers, two highly trained postdocs, and a new method for completing some lab procedure. And scientific progress can appear as another expected outcome on the list, measured by expert elicitation about how much the experimental findings would shift their priors about the hypothesis in question as discussed above.

Ideally, each outcome vector $x(s)$ would be assigned a utility $u(x(s))$ by the procedure sketched in (2c) above. In practice, a Likert scale of some cruder estimate should be quite sufficient. What a utility function must capture are attitudes towards trade-offs. Which is better, an outcome with three good papers and one well-trained postdocs or an outcome with two such papers and two such postdocs? One particular kind of trade-off encoded by a utility function concerns attitudes towards risk. The more risk averse you are, the more the expected utility $\frac{1}{2}u(v_1) + \frac{1}{2}u(v_2)$ of flipping a fair coin and getting either output vector $v_1$ or $v_2$ is exceeded by the utility $u(\frac{1}{2}v_1 + \frac{1}{2}v_2)$ of receiving their convex combination for sure.

[11] Solomon Kullback and Richard A. Leibler, "On Information and Sufficiency," Annals of Mathematical Statistics 22, no. 1 (1951): 79–86, https://doi.org/10.1214/aoms/1177729694.

Faced with trade-offs like this when making multi-objective decisions, people are often tempted to construct a utility function by taking a weighted average of how they would rate each objective independent of the others, i.e., by assuming a linear utility function of the form $u(v) = w \cdot v$ where $w$ is a weight vector. To be concrete, that might look like a weighted average of how many papers are promised and how many postdocs. Though common, this kind of procedure rarely has very good mathematical properties in complicated situations like this. It implies, for example, that trade-offs between one objective and another can be achieved at a constant rate. (For a better and more general approach to multi-objective decision-making, the Method of Equal Swaps prescribes imagining choices that don't exist[12] but that, because of transitivity, help clarify your preferences.)

The key point is that the utility of an output vector can, and should, be considered to reflect NIH priorities independent of any probability estimates. In other words, specifying a utility function is a matter of saying of saying how much better this outcome would be *if you had it for sure* compared to another outcome if you had that one for sure. Once we have estimates for both the utility $u(x(s_i))$ and the probability $p(s_i)$ for each state $s_i$ in $S$, all that is left to do is compute the expected utility $\sum[u(x(s_i)) \times p(s_i)]$ of option $x$ and select it over option $y$ if and only if the expected utility $\sum[u(y(s_i)) \times p(s_i)]$ is smaller.

4. <u>Theories of change</u>

As a step towards making our stylized account of grant selection a bit more realistic, note that NIH does not consider a handful of proposals at a time but rather receives well over 50,000 per year and funds just under 20% of those.[13] That makes it quite infeasible to imagine in detail all the various states of the world that determine how these proposals would turn out if funded. As a practical matter, it is important to group together proposals whose outcomes primarily depend not on the detailed state of the world but more simply on certain kinds of states (i.e., green or red) associated with certain broad hypotheses about the world. This is the role of a "theory of change."

---

[12] John S. Hammond, Ralph L. Keeney, and Howard Raiffa, "The Even-Swap Method for Multiple Objective Decisions," in Lecture Notes in Economics and Mathematical Systems (Berlin: Springer, 2000), https://doi.org/10.1007/978-3-642-57311-8_1; see also their Smart Choices: A Practical Guide to Making Better Decisions (Boston: Harvard Business School Press, 1999).

Specifically, a theory of change consists of three things. First, a partition of $S$ into disjoint subsets so that $S = S_1 \cup S_2 \cup \ldots \cup S_n$. Second, a probability function that assigns to each $S_i$ a nonnegative number $p(S_i)$ that represent the probability of Tyche's dart landing in $S_i$. And third, a particular kind of function $T: S \times A \to O$ that, given both a state of the world in $S$ and an action in $A$, determines an outcome such that, for each $x$ in $A$, the outcome $x(s)$ equals $x(r)$ when $s$ and $r$ both belong to the same partition component $S_i$. This condition is also captured by writing $x(S_i)$ in the context of a given theory of change, just as we did above for the probabilities.

For example, think of dividing the possible states of the world into two components: $S_1$ where the Warburg hypothesis about cancer is true, and $S_2$ in which it is not. A theory of change tells you what kinds of outcomes to expect from a given experiment $x$ depending on whether we live in a world where Warburg is true or not. Of course, as long as you assign nonzero probability to both components, such a simple theory of change can be refined over time to include many other possibilities besides a basic yes or no. That refined theory will then make use of a partition with more than two components. The partition encodes what you think matters about the states of the world that Tyche's dart determines.

So whereas *utilities* are about "what you want from the world" and *probabilities* are about "what you believe about possible worlds," a *theory of change* is about "what properties of the world determine how outcomes respond to actions." More simply, it is about under what circumstances one thing causes another. That is why a theory of change is often summarized by a causal graph. It is also why a good theory of change can help organize funding decisions. Here are three examples of how.

> **4a. Grouping proposals.** Which proposals should be compared with one another? Ideally, a proposal cohort should all belong to the set $A$ for a particular theory of change $T: S \times A \to O$. There are several reasons. First, the probability estimates needed to evaluate the expected utility of one option will be the same ones needed to evaluate the expected utility of another option, namely, the probabilities $p(S_i)$ of the components $S_i$ in the theory's partition of $S$. Second, when it comes to estimating utilities, the outcomes of one proposal in a given state of the world is apt to be more readily comparable to the outcome of another in the same state of the world because they all depend on the same theory of what matters. And third, the realized outcomes of these proposals will facilitate Bayesian updates of the original probability estimates if similar proposals continue to arrive. More specifically, you cannot really perform such updates unless the old and new proposals all belong to the set $A$ associated with a particular theory of change.

[13] See NIH Research Portfolio Online Reporting Tools (RePORT), "Success Rates," https://report.nih.gov/, and the NIH Data Book, https://report.nih.gov/nihdatabook/. For example, in FY2020 NIH received 55,038 competing research-project-grant applications and funded 11,332, a 20.6% success rate.

**4b. Portfolio statistics.** Now suppose, more realistically, that the decision is to fund a number of proposals from a given group instead of just one. This again is facilitated if all the proposals under consideration belong to the set $A$ associated with a particular theory of change. In particular, you cannot really talk about the correlation between, say proposal $x$ and proposal $y$ otherwise. That is important because, in general when assembling a portfolio under uncertainty, it is best to include options that are pairwise either uncorrelated or, even better, anti-correlated.[14] This is certainly true if the utility of a portfolio is quadratic in the sum of the output vectors since the variance matrix of that sum is the sum of the variance matrices plus covariance terms. Analyzing risk in these terms is everyday practice when evaluating financial portfolios, for example.

**4c. Encouraging proposals.** Sometimes funders do not simply reject a proposal but rather suggest ways of changing it into something more worthy of support. Again, that process is hardly makes sense unless the new and old versions both belong to the set $A$ associated with a particular theory of change. Funders also can try influencing the proposals the candidate proposals that can consider for support by issuing an "advanced market commitment."[15] In terms of a theory of change, that requires specifying a target subset of the outcome space $O$ and promising a reward to any researcher whose plan, when realized, achieve outcomes in that target subset. Note that such targets should not be chosen based only on the utility of its members. Developing a successful commitment to announce also requires a theory of change to check that achieving those target outcomes seems doable enough for researchers to want to try.

[14] Harry Markowitz, "Portfolio Selection," Journal of Finance 7, no. 1 (1952): 77–91, https://doi.org/10.1111/j.1540-6261.1952.tb01525.x.
[15] See Owen Barder, Michael Kremer, and Heidi Williams, "Advance Market Commitments: A Policy to Stimulate Investment in Vaccines for Neglected Diseases," The Economists' Voice 3, no. 3 (2006), https://doi.org/10.2202/1553-3832.1144; and Michael Kremer, Jonathan Levin, and Christopher M. Snyder, "Advance Market Commitments: Insights from Theory and Experience," AEA Papers and Proceedings 110 (2020): 269–273, https://doi.org/10.1257/pandp.20201017.

On this view, the science of science exists to help describe, clarify, and improve change theories, probability estimates, utility functions, and other elements of decision-making under uncertainty that necessarily play roles—explicitly or implicitly—in choices made by scientists and scientific funders. Note that this does not reduce those choices to objective technicalities. While the framework described helps organize and optimize the decision-making process, specifying elements such as a theory of change does depend ultimately on specific kinds of experience and judgement that can be sought out, explained, and even challenged much more meaningfully than just asking experts directly to rate, based on their opinion, how suitable various proposals are for funding.

## 5. How mature sciences make and measure progress

Which is more mature, natural science or this kind of science of science? Both natural scientists and metascientists are trying to make decisions under uncertainty. The laboratory scientist is concerned with which experiments to try running out all the possibilities. The metascientist is concerned about which proposals to fund out all the possibilities. Both have to take guesses, wait for Tyche to throw her dart, then update those guesses depending on the realized outcomes.

At least that is what a Bayesian does. The entrenched way that most natural scientist deal with uncertainty is different from this, relying on the frequentist assumption that the exact same experiment can be run many, many times. This seems obvious as long as the laws of nature are not changing. But experience shows at least three reasons for doubt. One is that, in any situation—like those studied by metascience—where human or societal or historical forces are work, the laws governing the system may well change. Another is that, despite pleas for researchers to conduct reproductions, very few ever take place.[16] Moreover, when same experiment is run again, an easy conclusion to draw in many cases is that it wasn't really the very same experiment to begin with. Or, in other words, that the causes at work were not well understood yet. More specifically:

---

[16] Open Science Collaboration, "Estimating the Reproducibility of Psychological Science," Science 349, no. 6251 (2015): aac4716, https://doi.org/10.1126/science.aac4716, which found that only about a third of a sample of 100 published psychology studies replicated.

5a. **Reproducibility and scientific progress**: Bayesians update their priors based on the results of a hypothesis test. They have no problem assigning a probability to a hypothesis (through a process, say, described in §2c above), but they just don't believe that *my* probability and *your* probability about that probability need be the same—even though, with enough information, we should converge to the true probability under the questionable assumption that there is a true probability at all. Frequentists, by contrast, go through a stylized exercise of accepting or rejecting the null hypothesis that is firmly entrenched but hard to justify based on first principles. Given the arbitrary and sometimes manipulable nature of the criteria for making that determination (see *p*-hacking),[17] even well-meaning and careful attempts at reproduction can be difficult to interpret. Arguably, what a mature science means by reproducibility instead has more to do with performing multiple, high-quality tests of a given hypothesis and measuring scientific progress by eliciting the degree to which expert's probabilities about that hypothesis are converging as a result.

5b. **Causal analysis**: As noted in §4, a theory of change ultimately encodes some causal analysis. Natural science has strong, traditional, and implicit notions about causality that cannot be expressed very formally in terms of equations and are impossible to derive from observed correlations alone—suggestive though these may be. After all, the correlation of A with B is the same as the correlation of B with A, but saying that A causes B is very different from saying that B causes A. It is therefore impossible to derive reliable causal conclusions from correlational evidence alone as everyone knows but many routinely overlook—particularly those who think that Large Learning Models can deliver new causal insights.

Arguably, a mature science should have highly developed ways of justifying causal inference. Economists, for example, use a toolkit of techniques ranging from regression discontinuity design (RDD) and differences in differences (DiD) to randomized control trials (RCTs) and instrumental variables (IVs) depending on what the situation allows.[18] Though formally equivalent, Pearl's theory of causal graphs[19] can offer more accessible and straightforward ways of not only deciding whether explicit assumptions justify causal inferences but also deriving correct formulae for those inferences when they do.  Policy questions are almost all causal (e.g., If we do this, will that happen?), so such inferences are an important matter.

---

[17] Joseph P. Simmons, Leif D. Nelson, and Uri Simonsohn, "False-Positive Psychology: Undisclosed Flexibility in Data Collection and Analysis Allows Presenting Anything as Significant," Psychological Science 22, no. 11 (2011): 1359–1366, https://doi.org/10.1177/0956797611417632.
[18] On this econometric toolkit, see Joshua D. Angrist and Jörn-Steffen Pischke, Mostly Harmless Econometrics: An Empiricist's Companion (Princeton, NJ: Princeton University Press, 2009), https://doi.org/10.1515/9781400829828.

5c.  **Data needs**:  While a research grant's final report should, in principle, describe the outputs $x(s)$, not to mention how the results update beliefs about the probabilities of certain states of the world in $S$, in practice it is very difficult to obtain such data in any comprehensive, systematic, comparable, or long-term fashion.  One of the few attempts is the U-Metrics project housed at the Institute for Research on Innovation and Science (IRIS) at the University of Michigan.[20]  By collecting and collating universities' HR and Accounting records about grant spending, then linking that to Census, publication, patent, business, and other data, U-Metrics is a powerful proof of concept that has already begun delivering remarkable insights about the science of science.  One secret to its success so far is the importance of tracking *people*.  One impediment is that participation in U-Metrics by universities is limited and unlikely to grow without a funded mandate from NIH and NSF.

The reason NSF and NIH originally supported U-Metrics, with some help along the way from the Sloan Foundation, was largely for the purpose of institutional accountability—specifically cataloguing the results of federal funding provided under the American Recovery and Reinvestment Act of 2009 (ARRA).  The idea was to try facilitating more than just more cooking, as Romer might say.  In fact, the initiative was originally called STAR METRICS, which stood for Science and Technology for America's Reinvestment: Measuring the EffecT of Research on Innovation, Competitiveness and Science.[21]

---

[19] Judea Pearl, "Causal Diagrams for Empirical Research," Biometrika 82, no. 4 (1995): 669–688, https://doi.org/10.1093/biomet/82.4.669.
[20] The flagship UMETRICS output is Nikolas Zolas, Nathan Goldschlag, Ron Jarmin, Paula Stephan, Jason Owen-Smith, Rebecca F. Rosen, Barbara McFadden Allen, et al., "Wrapping It Up in a Person: Examining Employment and Earnings Outcomes for Ph.D. Recipients," Science 350, no. 6266 (2015): 1367–1371, https://doi.org/10.1126/science.aac5949.
[21] Julia Lane, "Assessing the Impact of Science Funding," Science 324, no. 5932 (2009): 1273–1275, https://doi.org/10.1126/science.1175335, describing the STAR METRICS initiative and its use of federal grant data, including American Recovery and Reinvestment Act funding, to measure research outcomes.

In conclusion, the decision theoretic view taken here shows that, while funders should not necessarily be evaluated based on particular outcomes (see §1 above), the focus should instead be on processes that are designed—based on provable first principles— to realize a pattern of valuable outcomes. As we have seen, this comes to down to formulating and updating a theory of change. And that can only be done with systematic and comprehensive data collection of the sort we now know how to implement. The conclusion is that the science of science is a mature science that, given proper support and data, can rigorously deliver insights, innovations, and measurable improvements regarding the meta-ideas about the production and transmission of ideas that Paul Romer referred to as "better recipes."